\documentstyle[aps,pre,epsf,twocolumn]{revtex}
\long\def\wideabs#1{\twocolumn[\hsize\textwidth\columnwidth\hsize%
\csname @twocolumnfalse\endcsname #1 \vskip1pc]}

\def\begineq{\begin{equation}}
\def\endeq{\end{equation}}
\def\ie{{\it i.e.,}}
\def\fone{{f^{(1)}({\bf r},{\bf v},t)}}
\def\ftwo{{f^{(2)}({\bf r_1},{\bf v_1},{\bf r_2},{\bf v_2},t)}}

\begin{document}
\draft
\widetext

\title{Transport Coefficients for Granular Media from Molecular
Dynamics Simulations\protect\footnotemark[1]}
\author{C. Bizon, M. D. Shattuck, J. B. Swift and Harry L. Swinney}
\address{Center for Nonlinear Dynamics and Dept. of Physics, University of
Texas, Austin, TX 78712}
\date{\today}

\wideabs{
\maketitle

\begin{abstract}
Under many conditions, macroscopic grains flow like a fluid; kinetic theory predicts continuum 
equations of motion for this granular fluid.  In order to test the theory,
we perform event driven molecular simulations of a two-dimensional gas 
of inelastic hard disks, driven by contact with a heat bath. 
Even for strong dissipation, high densities, and small numbers of
particles,  we find that continuum theory describes
the system well.  With a bath that heats the gas homogeneously, strong velocity
correlations produce a slightly smaller energy loss due to inelastic collisions
than that predicted by kinetic theory.  With an inhomogeneous heat bath,
thermal or velocity gradients are induced.  Determination of the resulting
fluxes allows calculation of the thermal conductivity and shear viscosity,
which are compared to the predictions of granular kinetic theory, and which can
be used in continuum modeling of granular flows.  The shear viscosity is close
to the prediction of kinetic theory, while the thermal conductivity can be
overestimated by a factor of 2; in each case, transport is lowered with
increasing inelasticity.
\end{abstract}
}

\vspace{0.3in}

\section{Introduction}
\label{intro}
\footnotetext[1]{Submitted to PRE Feb. 1999.}
Collections of inelastically colliding macroscopic particles, when
driven sufficiently by an external force, exhibit fluid like
behavior such as flow~\cite{campbell90,jaeger96} and instabilities~\cite{melo94,debruyn98}.  Although only a small number of particles ($10^2-10^6$) are typically 
involved in a
flow, continuum methods are popular tools in modeling, because a century of 
fluid dynamics experience can be brought to bear on the problem.
With continuum equations velocity profiles and transfer
rates can be calculated, and stability analyses performed.
Often,
plausible continuum equations are simply posited, but these equations typically
contain unmeasurable free parameters.  A more rigorous approach 
derives continuum equations from the kinetic theory of dissipative gases~\cite{lun84,jenkins85,jenkins85a}.
In principle, a closed system of partial differential equations results, 
analogous to the Navier-Stokes equations, with all transport coefficients
given.

However, the continuum equations for granular media do not share the stature of their molecular fluid
analogs.
For both grains and molecules, the general
form of the equations is not in question, but the constitutive equations
relating fluxes of momentum and energy to gradients may differ from the simple
cases of Newton's viscosity law and Fourier's heat law.  The transport
coefficients of liquids are not routinely calculated from kinetic theory, but
are measured experimentally;  such measurements have not been carried out for
granular media, where the granular temperature (the kinetic energy associated
with the fluctuational velocities of particles)  is difficult to control and
difficult to measure.  In addition, the small number of particles and
the dissipative nature of the medium have led some researchers to the 
conclusion that continuum approaches are doomed to failure~\cite{kadanoff97}.

As a result, the central question remains open: {\emph{Can continuum
equations, derived from kinetic theory and supplemented by measurements, model
rapid granular flows to the same level that Navier-Stokes equations model the
flows of liquids?}}   Molecular dynamics simulations will play a crucial role
in answering this question, since the data they provide can be used
to quantitatively test the assumptions and the results of kinetic theory.

The main assumptions, which we will elucidate in Sec.~\ref{kinetictheory}, are:
\begin{enumerate}  
\item Single particle distribution functions
are nearly Boltzmann, 
\item  Molecular chaos --- Particle velocities are uncorrelated, and 
\item Particle positions are correlated in accord with Eq.~(\ref{CandS}), the
Carnahan and Starling relation for elastic particles.  
\end{enumerate}
The main results, also discussed further in Sec.~\ref{kinetictheory},  are: 
\begin{enumerate}
\item The equation of state (Eq.~(\ref{state})), 
\item The constitutive relations (Newton's stress law and Fourier's heat law --- Eqs.~(\ref{newtonslaw}) and~(\ref{fourierslaw})),
and 
\item The values of the shear viscosity $\mu$, the thermal conductivity $\kappa$, and the loss rate of granular temperature due to inelastic collisions, $\gamma$ (Eqs. ~(\ref{viscosityenskog}),~(\ref{conductivityenskog}), and ~(\ref{gamma0eq})).
\end {enumerate}

We will use 
molecular dynamics simulations to test these points, and to measure 
the transport coefficients to be used in continuum analyses of granular flows.
We have found that the simulations quantitatively reproduce experiments on standing waves
in oscillated granular media, producing the correct wave patterns and 
wavelengths~\cite{bizon98} and secondary instabilities~\cite{debruyn98}. 
With the numerical simulation, we not only have access to experimentally 
unmeasurable quantities~\cite{bizon98a}, but we also can study systems that are not experimentally
realizable, allowing us to test both the assumptions and results
of the granular kinetic theory.  Once the constitutive relations have been
tested and possibly enhanced through the use of particle simulations, 
direct comparison between continuum theory and laboratory experiment becomes possible.

Laboratory experiments of granular kinetic theory
have not yet proceeded beyond measuring the
single particle velocity distribution function~\cite{olafsen98,delour98,oger96}.
Simulations that make contact with kinetic theory have focussed 
mainly on the homogeneous cooling state, a time-dependent state that 
eventually becomes spatially inhomogeneous.  In those simulations, long-range
velocity
correlations develop~\cite{orza98}, and molecular chaos no longer holds.
A more sophisticated form of kinetic theory, ring kinetic theory, in which
correlations in particle velocity are accounted for, is required for a 
full
description~\cite{vannoije98}.  Finally, a number of simulations have been
performed on one-dimensional granular media in contact with a heat
bath~\cite{williams96,swift98,puglisi98} to produce a steady state.  
In one-dimension (1D), spatial~\cite{williams96} and velocity~\cite{swift98} correlations can develop, and the single particle velocity distributions may be
nongaussian~\cite{puglisi98}.   The present work extends these 
1D steady state and 2D decaying simulations
by examining 2D steady states with the goal of 
quantitatively testing
the assumptions and results of granular kinetic theory.

In Section~\ref{sims} we describe the molecular dynamics simulations and the
forcing methods.  Section~\ref{homogeneous} discusses
simulations of granular media in which  the heat bath is spatially
homogeneous.  In that section, we will check the assumptions about
the nature of the single particle distribution function and the
correlations of position and velocity, as well as measuring the
equation of state and $\gamma$.  
In Section~\ref{inhomogeneous} we turn to our main
purpose: allowing the heat bath to vary spatially so that steady state
inhomogeneous states may be induced and described.  From these
simulations we can study the constitutive relations, the shear viscosity,
and the thermal conductivity. Section~\ref{conclusion} contains concluding
remarks.

\section{Kinetic Theory}
\label{kinetictheory}

We begin with a brief review of the kinetic theory of granular media,
which differs only slightly from the kinetic
theory of elastic particles as presented in textbooks such as~\cite{chapman}.
The number of particles in a volume and velocity element $d{\bf r}d{\bf v}$
centered at position ${\bf r}$ and velocity ${\bf v}$ is given by $\fone d{\bf r}d{\bf v}$;
$\fone$ is called the single particle distribution function.  Continuum 
quantities are given as averages over $\fone$.  In particular, the number 
density, average velocity and granular temperature are defined respectively as
\begin{eqnarray}
n({\bf r},t) &\equiv& \int \fone d{\bf v},\\
{\bf v}_0({\bf r},t) &\equiv& {{1}\over{n}} \int \fone {\bf v} d{\bf v},\\
T({\bf r},t) &\equiv& {{1}\over{nD}}  \int \fone ({\bf v}-{\bf v}_0)^2 d{\bf v},
\end{eqnarray}
where $D$ is the number of dimensions.
Note that the granular temperature $T$ is not the thermodynamic temperature 
of the particles due to the random motions of their molecules, but the
analogous kinetic energy due to the random motions of the macroscopic 
particles themselves.

The evolution of $\fone$ depends on the joint probability distribution,
$\ftwo$; collisions of two particles change
the single particle distribution.  As in molecular kinetic theory,  
Boltzmann's assumption of molecular chaos is made, \ie\ velocities are
assumed to be uncorrelated, although for sufficiently dense media, 
positional correlations are allowed:
\begin{eqnarray}
\lefteqn{f^{(2)}({\bf r}_1,{\bf v}_1,{\bf r}_2,{\bf v}_2,t) = }
 \nonumber \\
 & &  g(\sigma,\nu) f^{(1)}({\bf r}_2-\sigma {\bf\hat{k}},{\bf v}_1,t) f^{(1)}({\bf r}_2,{\bf v}_2,t),
\label{gchaos}
\end{eqnarray}
where $\sigma$ is the particle radius, ${\bf\hat{k}}$ is a unit vector
pointing from the center of particle 1 to the center of particle 2, and
$\nu ={{1}\over{4}} n \pi \sigma^2$ is the solid fraction in 2D.  

The 
positional correlations are accounted for through $g(r,\nu)$, the
radial distribution function, which is defined as the probability of having
a pair of particles whose relative distance lies in the interval $r,r+dr$, 
normalized by the probability for an ideal gas.  This function, evaluated
at the point of contact $r=\sigma$, gives the increase in the probability
of collisions due to dense gas (excluded volume) corrections.  For elastic
hard disks, spatial correlations are described by the formula of Carnahan and Starling~\cite{carnahan69},
\begineq
G_{CS}(\nu) = {{\nu(16-7\nu)}\over{16(1-\nu)^2}},
\label{CandS}
\endeq
where
\begineq
G(\nu) \equiv \nu g(\sigma,\nu).
\label{Gg}
\endeq
Equation~(\ref{CandS}) works well for elastic particles with solid fractions below 0.675, where a phase transition
takes place~\cite{alder62}, and is often 
used in modeling granular media~\cite{jenkins85a}.  Equation~(\ref{Gg}) is the
definition of $G$ in terms of the (unknown) radial distribution function $g(r,\nu)$, evaluated at $r=\sigma$, while
Eq.~(\ref{CandS}) is a particular model for $G$, denoted by the subscript $CS$.

An unforced collection of molecules approaches a Boltzmann distribution,
\begineq
\fone=n (2 \pi T)^{-D/2} e^{-C^2 / 2T},
\endeq
where $C \equiv |{\bf v}-{\bf v}_0|$.
Away from equilibrium, the local distribution for elastic particles is nearly Boltzmann.
Granular media dissipate energy with each collision, so that the equilibrium
state of an unforced granular medium is that of no relative motion.
For grains, the single particle distribution function is simply assumed to be nearly 
a Boltzmann distribution. 

With these assumptions, and the additional assumption that the
coefficient of restitution $e$ is only slightly less than 1 
(particles are only slightly inelastic), equations for the continuum mass,
momentum, and energy can be derived for disks in two dimensions~\cite{jenkins85,jenkins85a}:

\begin{eqnarray}
{{\partial n}\over{\partial t}} + {\bf \nabla} \cdot (n{\bf v_o})&=&0\\
n{{\partial {\bf v_o}}\over{\partial t}} + n{\bf v_o}\cdot{\bf \nabla v_o} &=& -{\bf \nabla \cdot \underline P}\\
n{{\partial T}\over{\partial t}} + n{\bf v_o}\cdot{\bf \nabla} T &=& -{\bf \nabla} \cdot {\bf q} - {\bf \underline P}:{\bf \underline E}-\gamma,
\
\label{balance}
\end{eqnarray}
where $E_{ij}={{1}\over{2}}(\partial_i v_{oj} + \partial_j v_{oi})$ are the
elements of the symmetrized velocity gradient tensor ${\bf \underline E}$. The
constitutive relations for the pressure tensor ${\bf \underline P}$ and heat
flux ${\bf q}$ are
\begineq
{\bf \underline P} = (P - 2 \lambda {\rm Tr}{\bf \underline E}){\bf
\underline I}  - 2 \mu ({\bf \underline E} - ({\rm Tr}{\bf \underline
E}){\bf \underline I})
\label{newtonslaw}
\endeq
and
\begineq
{\bf q} = - \kappa {\bf \nabla} T,
\label{fourierslaw}
\endeq
where Tr denotes trace and  ${\bf \underline I}$ is the unit tensor.
The 2D equations close~\cite{jenkins85a} with the equation of state, which
is the ideal gas equation of state with a term that includes dense gas
and inelastic effects,
\begineq
P = (4/\pi\sigma^2) \nu T [1 + (1+e) G(\nu)]
\label{state},
\endeq
and the predicted values, denoted with a subscript $0$,  for the bulk viscosity, $\lambda$,
\begineq
\lambda_0 = {{8 \nu G(\nu)}\over{\pi \sigma}} \sqrt{{T}\over{\pi}}
\endeq
the shear viscosity, $\mu$,
\begineq
\mu_0 = {{\nu }\over{2 \sigma}} [{{1}\over{G(\nu)}} + 2 + (1+{{8}\over{\pi}})G(\nu)] \sqrt{{T}\over{\pi}}, 
\label{viscosityenskog}
\endeq
the thermal conductivity, $\kappa$,
\begineq
\kappa_0 = {{2 \nu }\over{\sigma}} [{{1}\over{G(\nu)}} + 3 + ({{9}\over{4}}+{{4}\over{\pi}})G(\nu)] \sqrt{{T}\over{\pi}} ,
\label{conductivityenskog}
\endeq
and the temperature loss rate per unit volume, $\gamma$
\begineq
\gamma_0={{16 \nu G(\nu)}\over{\sigma^3}}(1-e^2)\left({{T}\over{\pi}}\right) ^{3/2}.
\label{gamma0eq}
\endeq

Because of the assumption of near elasticity, the coefficient of restitution
enters only in the equation of state and in the expression for the temperature
loss due to inelastic collisions.  To this order, the thermal conductivity and
shear viscosity are the same as those given by the Enskog procedure for
elastic disks~\cite{gass70}.   

\section{Driven Granular Media Simulations}
\label{sims}

We perform event-driven simulations~\cite{lubachevsky91,marin93} of a
granular gas in a two-dimensional periodic box of side length $L =
52.6 \sigma$, in contact with a thermal bath that stochastically heats
particles throughout the volume.  Between collisions, particles travel
freely.  In our model the collisions are instantaneous and binary;
they conserve momentum and dissipate energy.  Particles are assumed to
be frictionless; particle friction can be incorporated into kinetic
theories~\cite{lun87,lun91}, and was included in the simulations of
oscillated granular media,~\cite{bizon98,bizon98a,debruyn98}, but
introduces complications that we wish to avoid for this study.

When particles collide, new velocities are calculated by reversing
the component of the relative particle velocity along the line joining particle centers 
and multiplying it by the coefficient of restitution $e$, which is between
0 and 1.  If $e$ is independent of collision velocity, a finite time 
singularity can occur in the collision frequency, a phenomenon known as
inelastic collapse~\cite{mcnamara92,mcnamara94}.  To avoid this 
simulation-ending occurrence in a convenient and natural way, we allow $e$
to vary as
\begineq
e(v_n) = \left \{\begin{array}{cc} 1 - B v_n ^ {\beta} &, v_n <
v_a \\ \epsilon &, v_n > v_a
\end{array}
\right.  ,
\label{restform}
\endeq
where $v_n$ is the component of relative velocity along the line joining
particle centers (normal to the contact surface),
$B = (1-\epsilon)(v_a)^{-\beta}$, $\beta=3/4$ and $\epsilon$ is a constant, 
chosen to be $0.7$.  In simulations of oscillated granular media, the
results are not sensitively dependent on $v_a$ or $\beta$, and $\epsilon=0.7$
produces good agreement with experiment~\cite{debruyn98,bizon98,bizon98a}.
 In addition to forestalling collapse, variation of
$e$ allows us to further probe granular kinetic theory, which assumes that
$e \approx 1$.  As $T$ varies, the relative collision velocity will vary; hence so will $e$, and the relative importance of
that assumption can be gauged.  
At a given temperature, a
distribution of collision velocities and a corresponding distribution of
coefficients of restitution occur.  Varying $T$ varies not only the average
value of $e$, but also the amount of deviation around that average.

The variation in $e$ gives rise to a velocity scale that is not present in the
elastic case.  All quantities given below are nondimensionalized with the
particle diameter $\sigma$ and the crossover velocity $v_a$ at which the
coefficient of restitution becomes a constant.  In particular, the granular
temperatures all scale with $v_a^2$. 

Because inelastic collisions remove energy from the system, we must constantly add energy to achieve any sort of steady state.  The situation is opposite that in simulations on nonequilibrium systems of elastic particles, where the constant energy input from the driving must be removed through an artificial means~\cite{evans}.
 
Stochastic heating is performed in one of three
ways: A) \emph {White Noise} --- random kicks 
$\delta {\bf v}$ are added to particles' velocities, B) \emph{Random Accelerations} --- particles accelerate between collisions, and C) \emph {Boltzmann Bath} 
--- particle
velocities are obliterated and and replaced with velocities chosen from a 
Gaussian
distribution.  We discuss the motivation and implementation of each in turn.

\subsection{White Noise}

Williams and Mackintosh~\cite{williams96} introduced white noise as
a thermal bath for dissipative granular media.  In their model, a random velocity is added to each particle's
velocity during each time step $\Delta t$.  The velocities added to each
particle are not correlated with one another, nor are they correlated with
the velocities added in the previous $\Delta t$.  This model has the 
advantage that the equation of motion for particles between collisions may be 
written down as the Langevin equation
\begineq
{{d^2x_i}\over{dt^2}} = \zeta_i,
\label{Langevin}
\endeq
where $x_i$ is the position of the $i$-th particle and $\zeta_i$ is a Gaussian
white noise term, \ie ~$\langle\zeta_i(t)\zeta_j(t')\rangle = 2 F \delta_{ij}
\delta(t-t')$.  The fact that the heating can be analytically expressed makes
its inclusion into kinetic theory possible.

This forcing is straightforward to include in our simulations.  Rather than
adding the random kicks to particles all at once, we kick $2r_k$ randomly
chosen particles every time there is a collision.  The number $r_k$, then,
represents the ratio between the rate of kicks and the rate of collisions.
If the kicks are totally random, the center of mass momentum of the system 
will fluctuate, but we desire that the heat bath can only change the
fluctuational velocity, not the mean.  To ensure that the mean velocity
remains fixed, we apply random kicks to $r_k$ particles:
\begineq
{\bf v}_i \rightarrow {\bf v}_i + |\delta {\bf v}| \hat{\bf r}_i, 1\le
i\le r_k.
\endeq
The kicks themselves are all of the same size, $|\delta \bf v|$, but the 
directions $\hat{\bf r}_i$ are randomly chosen.  Then, kicks in the opposite directions
are applied to another $r_k$ randomly chosen particles:
\begineq
{\bf v}_i \rightarrow {\bf v}_i - |\delta {\bf v}| \hat{\bf r}_{i-r_k}, r_k<i\le 2 r_k.
\endeq

Because each kick requires recalculation of the kicked particle's collision
list, we want to minimize the kick frequency.  Empirically, we find that
our results are independent of $r_k$ for $r_k \ge 1$ and $r_k(\delta{\bf v})^2$ 
constant.  This is not surprising, since the average length of $r_k$ 
randomly oriented kicks of length $\delta\bf v$ is $\sqrt{r_k}\delta v$,  
and only the total kick between collisions matters when collisions occur.
For this reason, we perform most of our simulations with $r_k = 1$, where
the collision rate equals the kick rate.

\subsection{Random Accelerations: The Air Table Model}
\label{accelerated}

While white noise forcing has the advantage that it can be incorporated into
the kinetic theory relatively easily, it has the disadvantage that it does
not model any particular real system.  In most experiments, energy is added
to the granular media through a boundary, causing gradients in  
the energy perpendicular to  that boundary.  For a vertically oscillating layer in a gravitational field~\cite{melo94}, the
situation is even more severe; although one might like to imagine the oscillating wall as thermal, 
providing stochastic kicks to the layer, the interactions of the layer with
the plate are strongly correlated to the plate's motion.  The plate introduces
both spatial and temporal gradients.  

One experimental system is capable of producing homogeneous steady states,
specifically, a collection of pucks on an air table~\cite{oger96,ippolito95}.
Although the purpose of the air is to levitate the pucks, it also 
drives their horizontal motion.  Either due to inhomogeneities in
the air flow or because the pucks are not perfectly parallel to the surface of
the table, the pucks accelerate uniformly from one collision to
another~\cite{oger96}.  This driving counters the loss of energy due to
inelastic collisions and produces a steady state.  

We model this air table driving by allowing each particle to move under a 
uniform acceleration: 
\begineq
{\bf a}_i = a_0 \hat{\bf r}_i
\endeq 
The magnitudes of all particle accelerations, $a_0$, are the same, but the directions, $\hat{\bf r}_i$, are randomly and uniformly chosen.  
The direction of a particle's acceleration changes stochastically.
When a collision occurs, $2r_k$ particles are given new $\hat{\bf r}_i$.
In order to conserve total momentum, we hold the total acceleration of the
particles at zero by giving exactly opposite accelerations to pairs of 
particles.   Initially, each particle is paired with another, and these are
given opposite accelerations.
 Later, when one particle is chosen and its acceleration randomized,
its partner particle is also given a new acceleration, opposite to the
first particle's.

Experiments suggest that particles accelerate uniformly from collision to 
collision, with most of the changes in acceleration happening at collisions~\cite{oger96}.
Therefore, $r_k=1$ is probably a relatively good model for the air table 
experiments.  As $r_k$ increases, the particles feel a constant acceleration
over a small temporal range.  In the limit that $r_k\rightarrow\infty$, the model
is the same as the white noise model, which is rate independent.  In practice,
we observe that the distribution of collision velocities for the accelerated
particle model approaches that for the white noise model at $r_k \approx 8$. 

Variation of coefficient of restitution with normal collision velocity is
critical for simulations with the accelerated forcing, since inelastic collapse-like collision sequences are more prevalent.  After a collision, particles 
move away from one another.  Without relative acceleration, they must collide
with other particles to reverse their velocity if they are to recollide.  
Particles with
relative acceleration, however, may collide and recollide without
striking another particle in the intervening time, just as a ball bounces
repeatedly on the ground.  On average,  the relative
acceleration is likely to change during this interval, but that is not guaranteed for any given pair.  Further, since accelerations are changed when collisions
occur, the rate of acceleration fluctuation is dependent on the collision rate
in the entire medium.  A given pair of particles can hijack the collision 
sequence of the gas, rapidly recolliding with one another.  If the coefficients
of restitution are constant, this scenario will produce collapse.  By allowing
the collisions to become more elastic for decreasing relative velocities, however, collapse is prevented; eventually, the relative acceleration will change, and the particles will move apart.

\subsection{The Boltzmann Bath}

Finally, we introduce a heat bath that approximates the assumption of
molecular chaos.  Molecular chaos assumes that the velocities of
colliding particles are uncorrelated; particles collide, and before
they collide with one another again, they collide with a large number
of other particles, losing the memory of the initial condition.  A
strong heat bath can perform the same function if it replaces the
particle velocities with new velocities chosen from a given
distribution.  If the heat bath interacts with the particles 
often enough, molecular chaos will be guaranteed.  While this bath is
wholly unphysical, it produces a situation in which the kinetic theory
is expected to apply exactly; it is a useful check on calculations,
and helps to elucidate the role of velocity correlations.

Implementation of the Boltzmann bath is simple.  When two particles collide,
$2r_k$ particles are randomly chosen and given velocities chosen from a Boltzmann
distribution with a specified temperature.

\section{Homogeneous Forcing, Correlations, and the Equation of State}
\label{homogeneous}

With any of the thermal baths just described, we can set up inhomogeneous states by varying the strength or rate of forcing over space.  However, in
the simplest case we force homogeneously.  The simulations are performed
for a variety of
solid fractions, forcing rates, and forcing strengths for the three thermal baths.

\begin{figure}
\epsfxsize=.9\columnwidth
\centerline{\epsffile{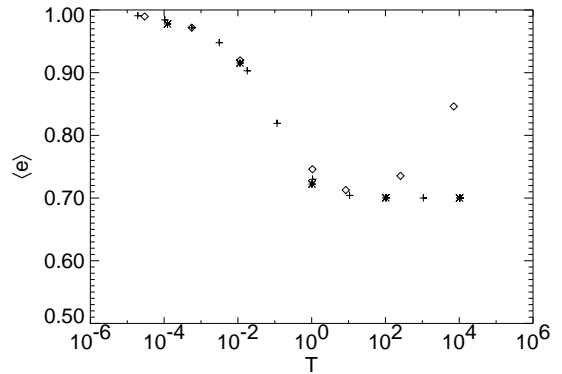}}
\smallskip
\caption[Average coefficient of restitution versus temperature] {Average coefficient of restitution vs T for $\nu=0.5$, for the three forcing methods. $+$: White noise forcing, $\diamond$: Accelerated forcing, $*$: Boltzmann Bath.  As
discussed in the text, $T$ is nondimensionalized with $v_a^2$.}
\label{restitution}
\end{figure}

The
average coefficients of restitution for a number of runs at different
temperatures are shown in Fig.~\ref{restitution}. 
Note that at high temperatures, the average coefficient of restitution begins
to rise for the accelerated forcing.   This is due to the inelastic collapse-like collision sequence described earlier for this forcing.  As the magnitude of acceleration increases to produce higher temperatures, these multiple collisions become
more and more important, leading to a large number of collisions with very
low velocities, and so a high average coefficient of restitution.

In the state produced by homogeneous forcing, we can measure the
single particle distribution function, the temperature produced by the forcing,
the pressure, the radial distribution function, velocity correlations, and loss
rates.  These quantities can be compared to the
corresponding quantities for elastic simulations and to their assumed or
calculated values from kinetic theory.

\subsection{Single Particle Velocity Distributions}

\begin{figure}
\epsfxsize=.9\columnwidth
\centerline{\epsffile{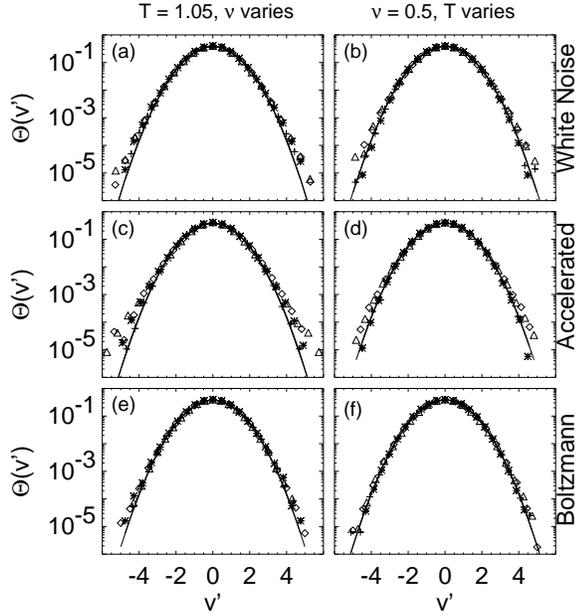}}
\smallskip \caption[Single particle distribution functions] {Single particle
distribution function $\Theta(v^{\prime})$ for (a) and (b) White noise forcing, (c) and (d)
Accelerated forcing, and (e) and (f) the Boltzmann bath, all with $r_k=1$.  The
velocities are scaled with the temperature $T$, so that $v^{\prime} = v/\sqrt{T}$ and
$\Theta(v^{\prime}) = Pr(v^{\prime}) \sqrt{T}$, where $Pr(v^{\prime})$ is the
probability distribution of $v^{\prime}$. In the left column, the average
temperature is approximately 1.05, and the solid fraction is varied ($+$:
$\nu=0.1$, $*$: $\nu=0.4$, $\diamond$: $\nu = 0.6$, $\triangle$: $\nu = 0.8$.)
In the right column, $\nu$ is fixed at $0.5$ and the temperature is varied;  (b)
$T= (+) 1.93 \times 10^{-5}, (*)3.13\times
10^{-2},(\diamond)1.06,(\triangle)1067.$ (d) $T = (+) 3.0 \times 10^{-5}
,(*)1.1\times 10^{-2},(\diamond)1.05,(\triangle)256.$ (f) $T = (+) 1.2\times
10^{-5},(*)1.1\times 10^{-2},(\diamond)1.02,(\triangle)102.$  The solid curves
are Boltzmann distributions.}
\label{oneparticledist}
\end{figure}

The lowest order approximation to $f$ in the kinetic theories is usually chosen to be a Boltzmann distribution,  the form of $f$ for an undriven elastic
gas.  A driven inelastic gas, although it approaches a steady state, is by
no means guaranteed to act like an undriven elastic gas.  Nevertheless,
the single particle distribution functions measured from the simulation
are all close to Boltzmann distributions, as seen in Fig.~\ref{oneparticledist}.

\begin{figure}
\epsfxsize=.9\columnwidth
\centerline{\epsffile{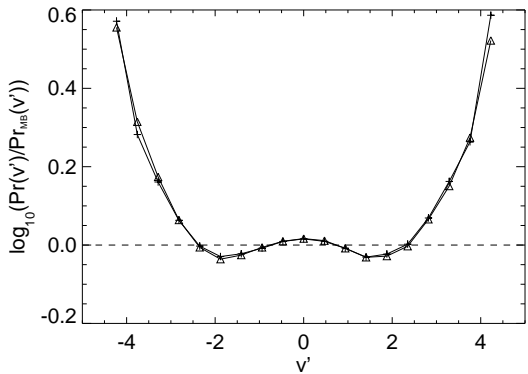}}
\smallskip
\caption[Deviations of velocity distribution from Gaussian] {The velocity
distribution function $Pr$ of $v^{\prime}=v/\sqrt{T}$ from a simulation with accelerated forcing at $\nu=0.5$ and
$T=1.05$, divided by $Pr_{MB}$, a Maxwell-Boltzmann distribution with $T=1.05$.
The two curves are for the two velocity components.}
\label{ratio}
\end{figure}

Overall, the accelerated forcing produces the strongest deviations from Maxwellian, and the Boltzmann bath, unsurprisingly, produces the least deviation.  In
all cases, the deviations become stronger as the density and temperature
increase (recall that increasing temperature has the same effect as decreasing the
average coefficient of restitution).  These deviations tend to flatten the
distribution, increasing the probability in the tails and slightly in the peak,
and decreasing the probability in between, as displayed in Fig~\ref{ratio}.
Similar types of deviations, but much stronger, have been observed in
experiments on a dilute, vertically
oscillated granular layer~\cite{olafsen98}.

\subsection{Equation of State and the Radial Distribution Function}

The equation of state,~(\ref{state}), relates the pressure, density and
temperature to the coefficient of restitution and $G(\nu)$.
The virial theorem of mechanics as applied to hard spheres can be
used to calculate the equation of state ~\cite{hirschfelder,rapaportsbook}.
\begineq
P V = N T + {{\sigma}\over{2 t_m}}\sum_{c} {\bf \hat{k}} \cdot \Delta{\bf v}_i,
\label{mdstate}
\endeq
where the sum is over all collisions that occur during the measurement time $t_m
$, $\Delta{\bf v}_i$ is the change in the velocity of the i-th particle
due to the collision, and $\bf \hat{k}$ is the unit vector pointing from particle
center to particle center.  In this form, measurement of
pressure reduces to measurement of the average particle energy and the
average change in the normal velocity at collision;  we measure
pressure with this method.

Using Eq.~(\ref{mdstate}) to measure pressure, and assuming the equation of state
(Eq.~(\ref{state})), we  produce a measurement of $G(\nu)$, denoted $G_{s}(\nu)$, where the subscript $s$ stands for simulation.  This measured value of $G$
will be compared to the Carnahan and Starling value $G_{CS}(\nu)$ from Eq.~(\ref{CandS}).
Accurate characterization of $G(\nu)$ is important, because it occurs 
in the expressions for transport coefficients. 

\begin{figure}[tb]
\epsfxsize=.9\columnwidth
\centerline{\epsffile{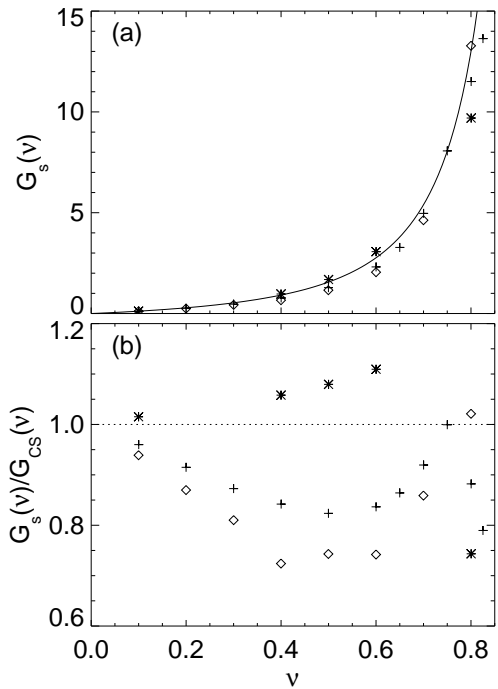}}
\smallskip
\caption[Equation of state for inelastic disks] {(a) $G_{s}(\nu)$ for inelastic hard discs driven by $+$: white noise forcing, $\diamond$: accelerations, and $*$: Boltzmann forcing. The solid curve is the Carnahan and Starling relation $G_{CS}(\nu)$, given by~(\protect{\ref{CandS}}). (b) The ratio of $G_s(\nu)$ to $G_{CS}(\nu)$.  All runs have $r_k=1$ and $T=1.05$.}
\label{Ginelastic}
\end{figure}

We calculate $G_s(\nu)$ for the three types of forcing as $\nu$ varies. The 
results are shown in Fig.~\ref{Ginelastic}.  For $\nu$ below $\approx 0.675$,
where elastic particles undergo a phase transition to an ordered 
state~\cite{alder62}, the white noise and accelerated runs 
produce lower $G$ than  elastic runs; Boltzmann runs have $G_s(\nu)$ 
slightly above the elastic values.  

\begin{figure}
\epsfxsize=.9\columnwidth
\centerline{\epsffile{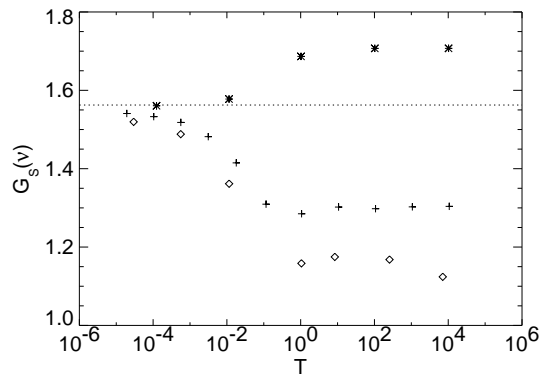}}
\smallskip
\caption{$G_s(\nu)$ vs T for $\nu=0.5$. $+$: white noise, $\diamond$: accelerated, $*$: Boltzmann. The dotted line is the Carnahan and Starling relation $G_{CS}(\nu)$, given by~(\ref{CandS}), for $\nu=0.5$.}
\label{GvsT}
\end{figure}

As the temperature decreases, 
$e\rightarrow 1$, and the values of $G_s(\nu)$ must approach the elastic 
values.  Therefore, $G_s(\nu)$ must be temperature dependent; this 
dependence is shown in Fig~\ref{GvsT}, along with the value of $G_s(\nu)$ 
given by Eq.~(\ref{CandS}).  As $T$ decreases, the inelastic $G_s(\nu)$ 
approaches the elastic $G$, and at high $T$, where $e$ is independent 
of $T$, $G_s(\nu)$ becomes independent of $T$.  As for the single particle 
distributions, the accelerated forcing shows the greatest deviation from 
the elastic behavior.

The substantial differences between $G_s(\nu)$, deduced from the equation of 
state, and $G_{CS}(\nu)$, predicted by the Carnahan and Starling relation (Eq.~(\ref{CandS})),
comes from two sources.  First, Eq.~(\ref{CandS}) may not give the correct value
of the radial distribution function at the point of contact, $g(\sigma,\nu)$, for granular media.  Second, $G_s(\nu)$ may not equal $\nu g(\sigma,\nu)$.  
Because Eq.~(\ref{Gg}) is the definition of $G$, this would amount to a change
in the equation of state, which we assumed to be correct in the calculation
of $G_s(\nu)$.

\begin{figure}
\epsfxsize=.9\columnwidth
\centerline{\epsffile{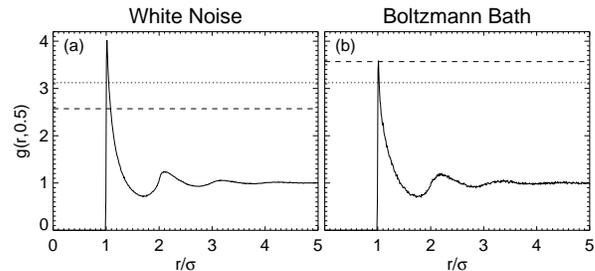}}
\smallskip
\caption[Radial distribution function for inelastic particles] {$g(r)$ at $\nu=0.5, T=1.05$ for (a) white noise ($r_k=1$) and (b) Boltzmann ($r_k=4$) forcing. The
dot-dashed lines represent the value of $g$ given by the Carnahan and
Starling relation Eq.~(\ref{CandS}) for $\nu=0.5$,
while the dashed lines show $G_s(\nu)/\nu$. For white noise forcing, $g(\sigma,\nu)$ coincides with neither line, while for the Boltzmann bath, $g(\sigma,\nu)$
coincides with $G_s(\nu)/\nu$.}
\label{gofrinelastic}
\end{figure}

To elucidate these two causes, we plot $g(r,\nu)$ for a run with white noise
forcing and a run forced with a Boltzmann bath.  Each plot also indicates
the value of $g(\sigma,\nu)$ predicted by Eq.~(\ref{CandS}), as well as that
predicted by Eq.~(\ref{Gg}), assuming $G(\nu)=G_s(\nu)$.  For neither forcing
type does~(\ref{CandS}), the Carnahan and Starling relation for $G_{CS}(\nu)$, properly predict $g(\sigma,\nu)$;  rather, 
inelastic particles are more likely to be nearly in contact than elastic particles
at the same density and temperature.  Furthermore, while $G_s(\nu) = \nu g(\sigma,\nu)$ for the Boltzmann forcing, this does not hold for the white noise
forcing or the accelerated forcing (not shown).  Even though $g(\sigma,\nu)$
is larger than that predicted by the Carnahan and Starling relation, $G_s(\nu) < G_{CS}(\nu)$, indicating that the equation of state is incomplete.
  Recalling that the Boltzmann
driving represents particles in contact with a highly  randomizing bath,
we conclude that the failure of the equation of state, (\ref{state}), is due to incomplete randomization
of particle velocities through collisions, or in short, a breakdown of
molecular chaos. 

\subsection{Velocity Correlations}

Molecular chaos is the assumption
that particle velocities are uncorrelated.  Knowing the velocity of one of
a pair of a colliding particle gives no information about the velocity of
the other.   In light of the behavior of $G$, and simulations of driven 1D 
and cooling 2D gases that
showed strong velocity correlations~\cite{swift98}, we measure velocity-velocity
correlation functions.

Given two particles, labeled $1$ and $2$, ${\bf \hat{k}}$ is a unit vector
pointing from the center of $1$ to the center of $2$.  Particle $1$'s velocity
then has a components $v_{1}^{||}$ parallel to and $v_{1}^{\perp}$ perpendicular to ${\bf \hat{k}}$; likewise for particle $2$.  We define two correlation functions
\begin{eqnarray}
\langle v_{1}^{||} v_{2}^{||} \rangle &=& \sum v_{1}^{||} v_{2}^{||} / N_r,\\
\langle v_{1}^{\perp} v_{2}^{\perp} \rangle &=& \sum v_{1}^{\perp} v_{2}^{\perp} / N_r,
\end{eqnarray}
where the sums are over $N_r$ particles such that the distance between
the two particles is within $\delta r$ of $r$.  If particle
velocities are uncorrelated, $\langle v_{1}^{||} v_{2}^{||} \rangle$ and $\langle v_{1}^{\perp} v_{1}^{\perp} \rangle$ will both give zero.

\begin{figure}
\epsfxsize=.9\columnwidth
\centerline{\epsffile{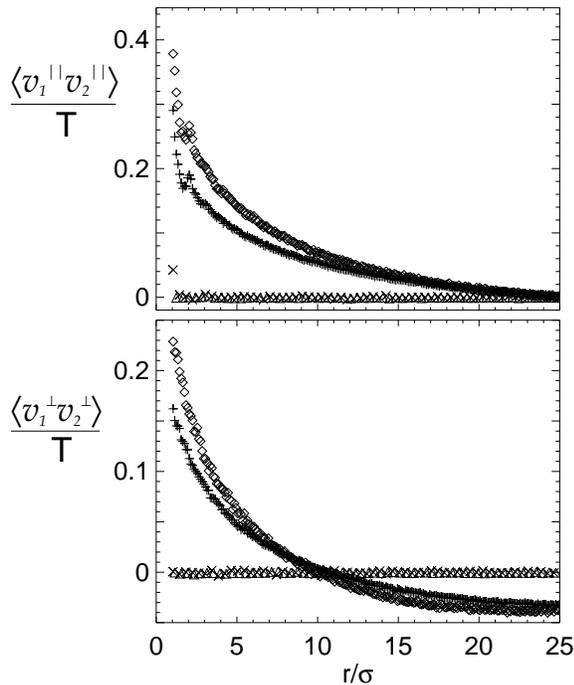}}
\smallskip
\caption[Velocity correlation functions]{Velocity correlations as a function
of particle separation at $\nu = 0.5, T=1.05$ for: ($+$) white noise,
($\diamond$) accelerations, ($\times$) Boltzmann forcing, and ($\triangle$)
elastic particles.  Each curve is built from around 100 frames separated in
time by 100 collisions per particle, and $\delta r = \sigma/10$.  Both the
elastic particles and the particles forced with the Boltzmann bath have essentially zero correlation over most of the range.  The Boltzmann bath shows positive
correlations only at very short range.} 
\label{vllvpp}
\end{figure}

The parallel and perpendicular velocity correlations are plotted in Fig.~\ref{vllvpp} for the three types of forcing and for elastic particles.  Both for
particles driven with white noise and accelerations, strong long-range velocity
correlations are apparent, with more correlations produced by the accelerated
forcing, consistent with its stronger deviations in the single particle
velocity distribution and in $G$.  These correlations are not small, reaching
as much as $40\%$ of the temperature; typically, the perpendicular correlations
are about one-half of the parallel correlations.
Further, these correlations are long range
--- they extend the full length of the system.  The parallel correlations drop
to zero at L/2, while the perpendicular correlations reach zero around
$r=10\sigma$, and have a negative value but zero derivative at $L/2$.
The long-range of nature of the correlation is not due to the size of the
computational cell.  Similar cell-filling correlations were observed in runs 
4, 16, and 64 times larger~\cite{bizon99a}.

For the Boltzmann forcing, some
correlations are visible at very short range;  inelastic collisions are trying
to establish correlations, but before these correlations can be extended to
long range they are wiped out by the thermal bath.

\subsection{Loss Rate}

\begin{figure}
\epsfxsize=.9\columnwidth
\centerline{\epsffile{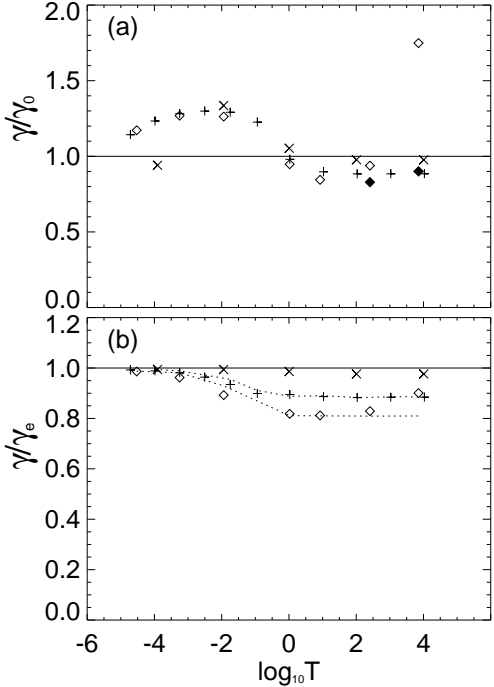}}
\smallskip
\caption{Temperature dependence of loss rate with $\nu=0.5$. (a) $\gamma/\gamma_0$, where the kinetic theory result $\gamma_0$ [Eq.~(\protect{\ref{gamma0eq}})] assumes a velocity-independent restitution coefficient. (b) $\gamma/\gamma_e$, where $\gamma_e$ takes into account the velocity dependence of $e$.  The symbols denote the type of forcing: $+$ White noise; $\diamond$ Accelerated; $\times$ Boltzmann. 
The dotted lines show ${{\sqrt{\pi}}\over{4\sqrt{T}}} \langle v_n^2\rangle_c / \langle v_n \rangle_c$ for the
white noise and accelerated runs; see (\protect\ref{avggamma}).} 
\label{gammaTpic}
\end{figure}

\begin{figure}
\epsfxsize=.9\columnwidth
\centerline{\epsffile{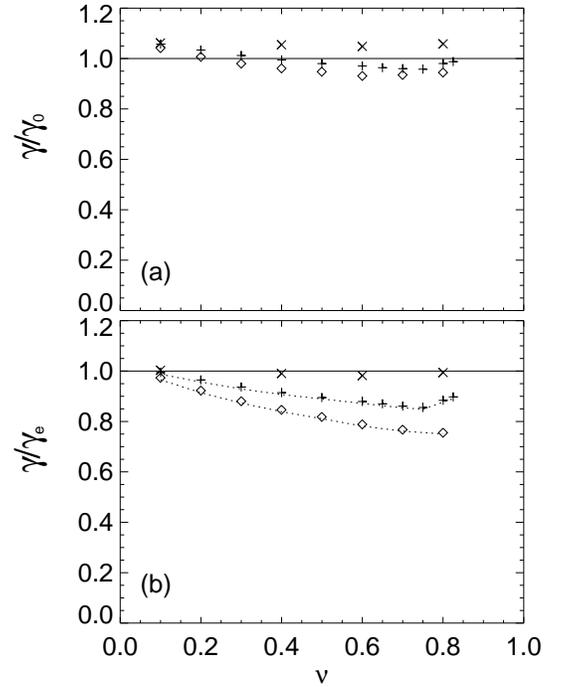}}
\smallskip
\caption{(a) $\gamma/\gamma_0$ and (b) $\gamma/\gamma_e$ versus $\nu$ at $T=1.05$ The symbols denote the type of forcing: $+$ White noise; $\diamond$ Accelerated; $\times$ Boltzmann} 
\label{gammanupic}
\end{figure}

The loss rate of temperature due to inelastic collisions, $\gamma$, divided by
the rate calculated from kinetic theory, $\gamma_0$ (see Eq.~(\ref{gamma0eq})), is
shown for the three forcing methods as a functions of 
$T$ in Fig.~\ref{gammaTpic}(a) and $\nu$ in Fig.~\ref{gammanupic}(a).   For the calculation of $\gamma_0$, $G$ was taken from the
equation of state measurements, and the average $e$ within a run was used.
Most surprising is the increased loss rate over kinetic theory at relatively 
low temperature.

The calculation of $\gamma$ requires only the evaluation of
\begineq
\gamma = {{\sigma}\over{2}}\int\int {{1}\over{4}}(1-e^2) ({\bf v}_{12}\cdot \hat{\bf k})^3 f^{(2)}({\bf v}_1,{\bf v}_2) d\Omega d{\bf v}_1 d{\bf v}_2,
\label{cintegral}
\endeq
where $d\Omega$ is the angle element. 
This expression simply averages the energy lost per collision,
\begineq
{{1}\over{4}}(1-e^2) v_n^2 = {{1}\over{4}}(1-e^2) ({\bf v}_{12}\cdot \hat
{\bf k})^2
\endeq
over all possible collisions.  The remaining factor of $({\bf v}_{12}\cdot \hat {\bf k})$ takes into account the fact that
grains traveling towards one another more rapidly are more likely to collide during any given time interval.

The kinetic theory result for the loss rate, Eq.~(\ref{gamma0eq}), follows from the collisional integral, Eq.~(\ref{cintegral}), under two assumptions: 
molecular chaos, which ought to be a more reasonable assumption at lower
temperatures, and the independence of $e$ on the variables of integration, so that it is pulled out of the integral as a constant.  At lower temperature,
where the variation of $e$ with $v_n$ leads to a distribution of $e$ at a 
given temperature, the second assumption fails.  As temperature drops still 
farther, the spread of $e$ reduces, since $e$ is bounded from above by $1$; at
the very lowest $T$, $\gamma$ does approach $\gamma_0$.

Substituting the function form of $e(v_n)$, Eq.~(\ref{restform}), into the collisional integral, Eq.~(\ref{cintegral}), assuming molecular chaos,  and performing the integrations, we arrive at an equation for $\gamma = \gamma_e$ that takes into account the variation of $e$ with $v_n$:
\begineq
\gamma_e={{4 \nu G \sqrt{T}}\over{\sigma^3 \pi^{3/2}}}\left[(1-e_0^2)(v_a^2+4T)
\exp(-v_a^2/4T) + 4 I\right],
\endeq
where
\begin{eqnarray}
I &=& 2^{1+\beta} A T^{1+\beta/2} (\Gamma(2+\frac{1}{2}\beta)-\Gamma(2+\frac{1}{2}\beta,v_a^2/4T))\nonumber\\ &-& A^2 2^{2\beta} T^{1+\beta}(\Gamma(2+\beta)-\Gamma(2+\beta,v_a^2/4T)),
\end{eqnarray}
$\Gamma(a)$ is the gamma function, and $\Gamma(a,b)$ is the incomplete gamma function.
In the limit that $v_a \rightarrow 0$, $\gamma_e \rightarrow \gamma_0$.

\begin{figure}
\epsfxsize=.9\columnwidth
\centerline{\epsffile{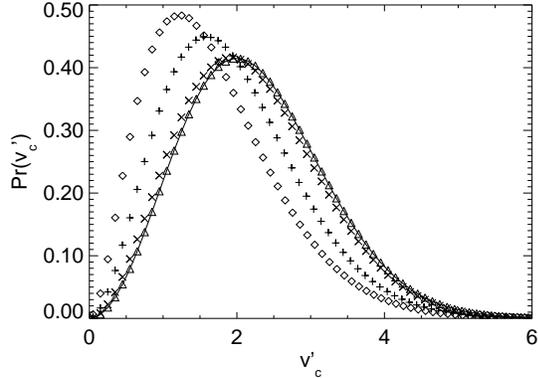}}
\smallskip
\caption[Probability distribution of relative velocities at collision]{Probability distribution of $v_c \equiv |{\bf v}_1 - {\bf v}_2|$, the magnitude of relative velocities at collision for different forcings
($+$) white noise, ($\diamond$) acceleration, ($\times$) Boltzmann, and ($\triangle$) for elastic particles, all at $\nu=0.5$ and $T=1.05$. The solid curve is the 
distribution predicted by uncorrelated collisions between particles chosen
from Boltzmann distributions, $(1/2\sqrt{\pi T^3}) v_c^2 e^{-v_c^2/4T}$. Positive
velocity correlation of nearby particles causes a reduction in the collision
velocities, and hence a reduction in $\gamma$.}
\label{PvsType}
\end{figure}

Figs.~\ref{gammaTpic}(b) and ~\ref{gammanupic}(b) show $\gamma/\gamma_e$ for the same simulations shown
in Figs.~\ref{gammaTpic}(a) and ~\ref{gammanupic}(a).  Taking the variations in $e$ into account
removes the underestimation of $\gamma$.  In the revised picture, $\gamma$
approaches $\gamma_e$ at low $T$ and at low $\nu$, but as either increases,
$\gamma$ drops from the value predicted by $\gamma_e$.
This is due to the velocity correlations produced by the inelasticity.
  Locally, particles are
moving together, reducing collision velocities and collision frequencies,
thereby reducing the loss rate; see Fig~\ref{PvsType}.  For the Boltzmann
forcing, velocity correlations are wiped out, and $\gamma$ is close to
$\gamma_e$.

Writing $\gamma$ in terms of average quantities makes its dependence on the
collision velocity more explicit.  The loss rate is identically equal to
the average energy lost per collision, $\langle \Delta E \rangle_c$, times the average collision frequency
per volume $f/V$
\begineq
\gamma=\langle \Delta E \rangle_c f/V = {{1}\over{4}} (1-e^2)\langle
 v_n^2\rangle_c f/V, 
\label{name}
\endeq
where the c subscript denotes an average over collisions, and assuming again 
that $e$ is independent of collision velocity.  
Similarly, the virial equation of state, Eq.~(\ref{mdstate}), in terms of average
quantities is
\begineq
P=(4/\pi\sigma^2) \nu T(1+{{V\sigma}\over{d}}f{{1+e}\over{2}}\langle v_n \rangle_c).
\label{avgG}
\endeq
Solving for $f$ in terms of $G$ from Eqs.~(\ref{state}) and ~(\ref{avgG}), and
substituting this into Eq.~(\ref{name}) we obtain
\begineq
\gamma={{(1-e^2)G}\over{\sigma}} {{\langle v_n^2\rangle_c}\over{\langle v_n\rangle_c}} nT.
\label{avggamma}
\endeq
If the distribution of relative normal velocity at collision is equal to that
predicted by molecular chaos, $P(v_n) = (1/2T) v_n \exp{-v_n^2/4T}$, then
$\langle v_n^2\rangle_c = 4T$ and $\langle v_n\rangle_c = \sqrt{\pi T}$, so
that $\gamma = \gamma_0$ is recovered.  

As seen in Fig.~\ref{PvsType}, however,
the distribution of collision velocities is different from the molecular chaos
values due to the 
presence of velocity correlations.  To show that this deviation accounts for
the remaining difference between $\gamma$ and $\gamma_e$,  $\langle v_n^2\rangle_c$ and $\langle v_n\rangle_c$ were calculated in the simulations.  Their
ratio, normalized by the molecular chaos value $4 \sqrt{T/\pi}$ is plotted
on Figs.~\ref{gammaTpic}(b) and ~\ref{gammanupic}(b) as dotted lines.  Except where there are wide
variations of $e$ within a given run (such as at high velocity in the accelerated runs), the change in the relative collision velocities tracks the change
in $\gamma$.

\section{Inhomogeneous Forcing and Transport Coefficients}
\label{inhomogeneous}

So far, we have been concerned with homogeneous forcing.  By 
using applied forcing that varies spatially, we can
induce inhomogeneous steady states.  Then, by measuring fluxes, 
transport coefficients are calculated, and compared to kinetic theory.
Inhomogeneous states have only been calculated for the accelerated
forcings; measurements for the homogeneous state show that deviations from
the elastic case are strongest in this case.

\subsection{Thermal Conductivity}
\label{thermalconductivity}

Recall that with the accelerated forcing, the direction of the accelerations
of particles fluctuated at a fixed rate, but the magnitude was always the
same.  To induce a thermal gradient in the simulation, we allow the magnitude
of the acceleration to vary in space.  Specifically, when the acceleration
of a particle is to be rotated, the magnitude of its acceleration is given
by 
\begineq
a_0(1-|(z-\frac{1}{2}L)/(\frac{1}{2}L)|),
\label{gradforcing}
\endeq
\ie we apply a linear gradient in the forcing, dependent
on one spatial direction ($z$), peaked in the center of the cell and falling to zero at the periodic boundary.   In order to preserve the center of mass 
momentum, the partner particle receives the opposite acceleration, regardless
of its position in the cell, as described in Section~\ref{accelerated}.

\begin{figure}
\epsfxsize=.9\columnwidth
\centerline{\epsffile{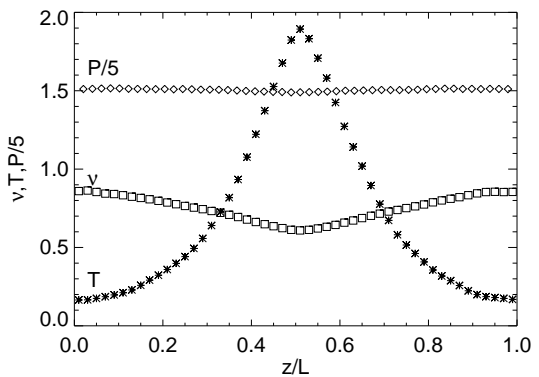}}
\smallskip
\caption[Temperature, density, and pressure profiles for inhomogeneous forcing]{
Profiles of $*$: T, $\Box$: $\nu$, and $\diamond$: $P/5$ for inhomogeneous forcing
in which the magnitude of the acceleration depends linearly on the distance 
from $z/L = \frac{1}{2}$.  The pressure is nearly constant, so that the variation in
$T$ induces a variation in $\nu$ through the equation of state. For this run,
the average solid fraction is $0.75$.} 
\label{profilesK}
\end{figure}

Under this forcing, a stable thermal profile develops, as seen in 
Fig.~\ref{profilesK}.   The system reaches a mechanical equilibrium; the
pressure is nearly constant in $z$.  Constant $P$ and varying $T$ imply 
varying $\nu$ through the equation of state, and this variation is observed.

\begin{figure}
\epsfxsize=.9\columnwidth
\centerline{\epsffile{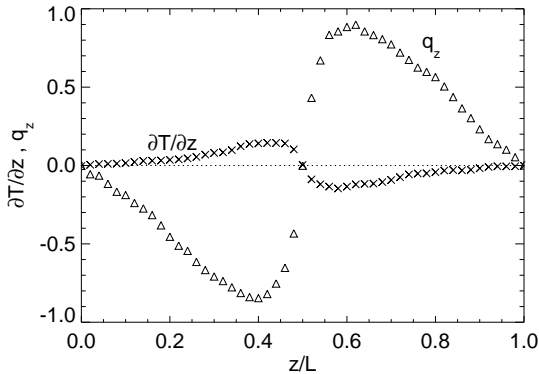}}
\smallskip
\caption[Heat flux and temperature gradient]{The $\times:$ thermal gradient and
$\triangle$: heat flux for the run shown in Fig.~\protect{\ref{profilesK}}.} 
\label{flux}
\end{figure}

The cell is divided into 50 slabs along $z$ for measurement purposes;  for each slab, 
snapshots allow calculation of the average $T$ and $\nu$ within.  Because
the pressure may in principle vary over space, its measurement  using the
virial fails.  Pressure is measured at the interfaces between the slabs
by keeping track of the normal ($z$) momentum flux through these boundaries, both
due to particles freely traveling through them, and due to collisions between
particles that lie in different slabs.  In addition, the energy added 
due to the forcing and the energy lost due to inelastic collisions are
separately accounted for each slab.   The difference between the
energy gain and loss in a given slab must, in a steady state, be made up
for by the difference in energy flux through its two boundaries, so that
the net rate of change of the energy in the slab is zero.  Assuming that the
energy flux through the line at $z=(0,L)$ is zero due to symmetry allows
calculation of ${\bf q}_z$, the heat flux through each slab boundary; see
Fig~\ref{flux}.

\begin{figure}
\epsfxsize=.9\columnwidth
\centerline{\epsffile{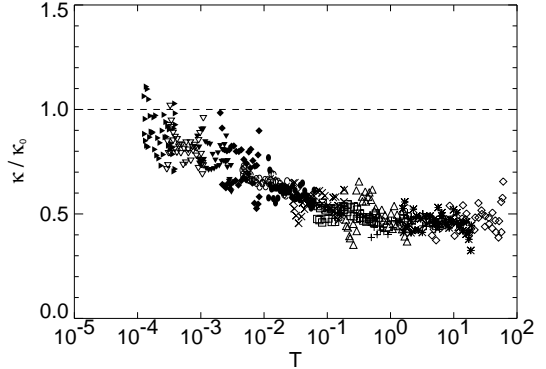}}
\smallskip
\caption[Deviation from Enskog theory for thermal conductivity]{Ratio of
$\kappa$ measured from simulations to $\kappa_0$ from Enskog theory (Eq.~(\ref{conductivityenskog})).  Each symbol 
denotes a different run, but for each run, the average solid fraction is $0.75$.} 
\label{Conductivity}
\end{figure}

Once ${\bf q}_z$ and $\partial T/\partial z$ are calculated, Fourier's heat law, Eq.~(\ref{fourierslaw}),
can be used to calculate the thermal conductivity $\kappa$. The results
of many simulations, holding the average $\nu$ at $0.75$, but varying 
$a_0$ in Eq.~(\ref{gradforcing}) and therefore the size of the thermal gradient,
are shown in Fig.~\ref{Conductivity}, are compared to the result of Enskog
theory, as given by Eq.~(\ref{conductivityenskog}).  At low temperatures, Enskog
theory does a good job predicting $\kappa$.  However, as the temperature
increases and $e(v_n)$ decreases, Enskog theory does worse; at the highest
temperatures, Enskog theory overestimates the thermal conductivity by a factor of 2.

 Note that this calculation does not test Fourier's Law; rather we assume
that Fourier's law is correct, and use it to calculate $\kappa$.  
Analysis based on closures of the Boltzmann
equation predict a term in the heat flux proportional to the density 
gradient~\cite{brey97a}.  If such a term had a sizeable magnitude and were
ignored, it would cause a reduction in the observed $\kappa$.



\subsection{Shear Viscosity}

Spatial inhomogeneity in the magnitude of the forcing led to a stationary
inhomogeneous temperature field, allowing measurement of heat flux and 
thermal conductivity; spatial inhomogeneity in the mean of the
forcing leads to a stationary inhomogeneous velocity field, allowing
measurement of the momentum flux and the shear viscosity.  In particular,
particle accelerations are chosen according to 
\begin{eqnarray}
a_y &=& a_0(0.01 \sin{(2\pi z/L)} +  \psi_0)\\
a_z &=& a_0 \psi_1
\end{eqnarray}
where $\psi_0$ and $\psi_1$ are numbers chosen randomly from a Gaussian distribution 
with zero mean and standard deviation of 1. 

\begin{figure}
\epsfxsize=.9\columnwidth
\centerline{\epsffile{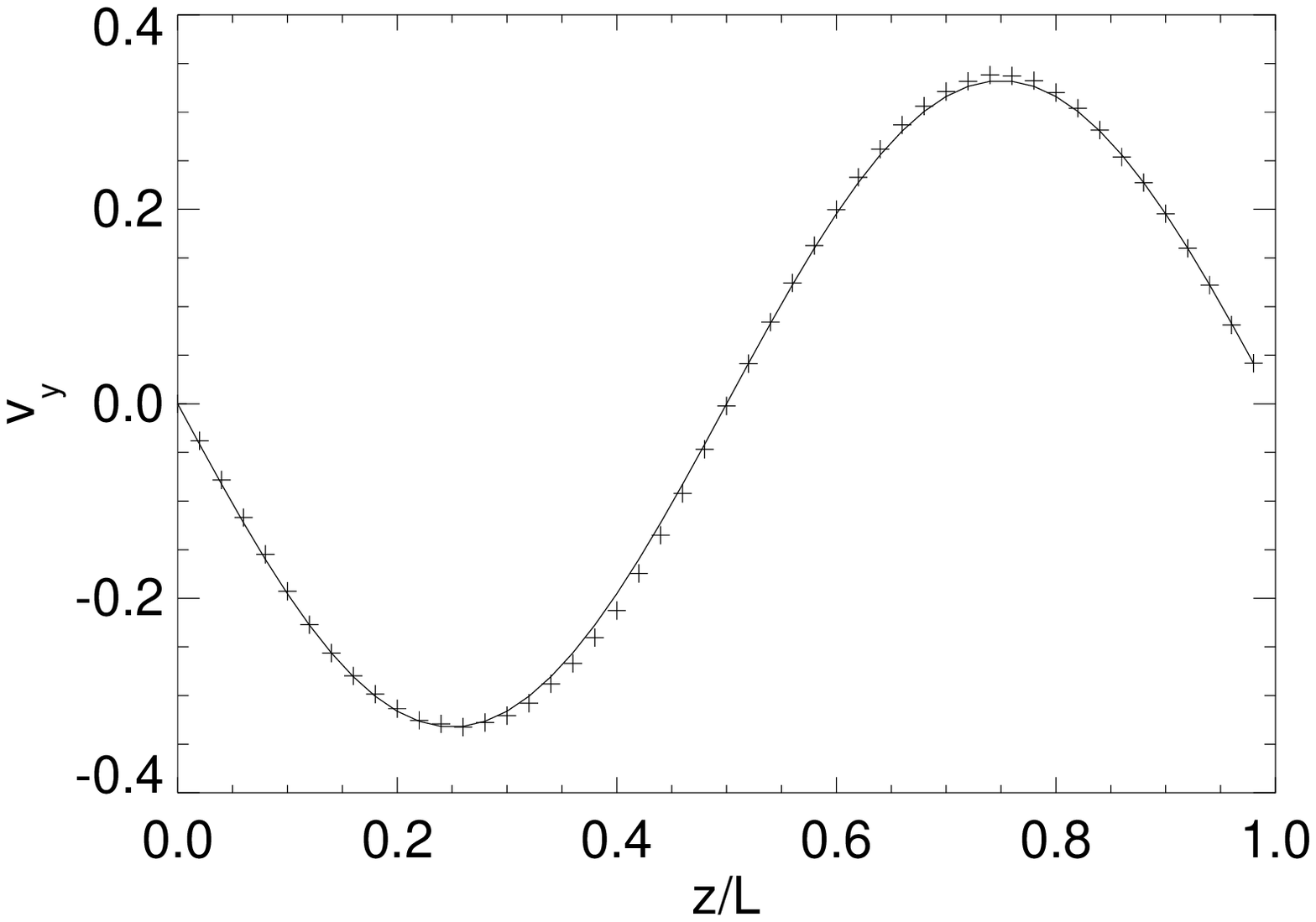}}
\smallskip
\caption[Velocity profile]{Average velocity in the $y$ direction as a function of $z$ at $T=0.21$, $\nu=0.6$. The solid line is sinusoidal.} 
\label{velocity}
\end{figure}


This forcing produces steady states with velocity, temperature, density, and stress fields like that shown in Fig~\ref{velocity}.
The velocity profile is nearly sinusoidal, and the
temperature, pressure, and solid fraction are essentially independent of $z$.  
In 
the simulations discussed so far, we have only considered the scalar quantity
$T = \langle (v - \langle v \rangle )^2\rangle / D$, where $D$ is the number of dimensions and the $\langle\rangle$ denote averages over particles.  This temperature
is more generally the trace of the temperature tensor:
\begineq
T_{ij}=\langle (v_{i} - \langle v_i \rangle )(v_{j}-\langle v_j  \rangle)\rangle,
\endeq
where $i,j$ range over the directions, and $v_i$ denotes the $i$-th component 
of the velocity.  In principle, $T_{yy}$ need not
equal $T_{zz}$ if the rate at which fluctuational energy is traded between
the directions is slower than the rate at which it is added anisotropically;
such is the case in a vertically  oscillated granular layer, where vertical fluctuational
energy can be twice that of the horizontal.
Introducing a bias in the acceleration of only 1\%, however, does not introduce
anisotropy into $T$; for larger shear rates this is not the case.  To study
the simplest case, we restrict our simulations to biases of 1\%.  

\begin{figure}
\epsfxsize=.9\columnwidth
\centerline{\epsffile{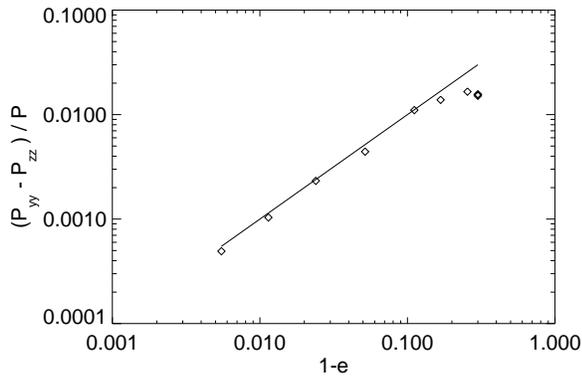}}
\smallskip
\caption[Stress difference vs. e]{Normal Stress difference divided by pressure
as a function of $1-e$.  The line has slope 1.} 
\label{stressdifference}
\end{figure}

In section~\ref{thermalconductivity} we described calculation of the pressure
by measuring the normal momentum flux through planes.  Measuring the
tangential flux through the planes, and introducing a second set of planes
orthogonal to the first, allows calculation of the full 4-component pressure
tensor.  As seen in Fig.~\ref{stressdifference}, the pressure tensor is anisotropic even though $T$ is isotropic;
the anisotropy increases as $T$ increases, or $e$ decreases.  For $e\approx 1$,
the stress difference is approximately proportional to $1-e$, but the variation
in $e$ within a given run probably plays a role, as it did in the loss rate; see
Fig.~\ref{stressdifference}.

\begin{figure}
\epsfxsize=.9\columnwidth
\centerline{\epsffile{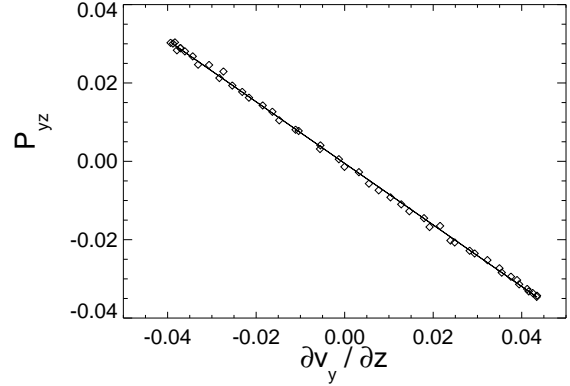}}
\smallskip
\caption[Stress versus strain rate]{The $yz$ component of the pressure tensor
versus the corresponding velocity derivative for the run of Fig.~\ref{velocity}.  The slope of the best fit line is $-\mu$.} 
\label{newton}
\end{figure}

\begin{figure}
\epsfxsize=.9\columnwidth
\centerline{\epsffile{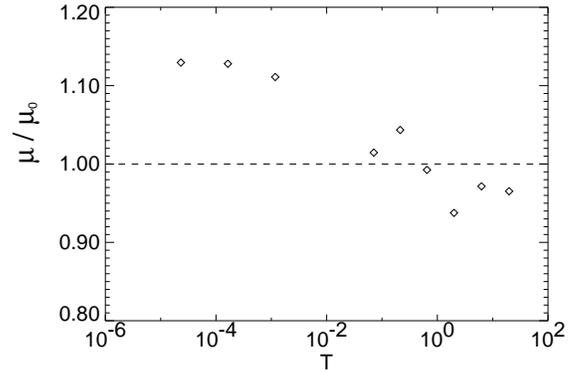}}
\smallskip
\caption[Viscosity compared to the Enskog values]{Viscosity, normalized by
the Enskog value $\mu_0$, as a function of T, for $\nu=0.6$.} 
\label{Viscosity}
\end{figure}

For each run at fixed $T$, we can test Newton's viscosity law,
\begineq
P_{yz} = - \mu {{\partial v_y}\over{\partial z}}
\label{viscouslaw}
\endeq
where the viscosity $\mu$ is a constant of proportionality.  A typical result
is shown in Fig.~\ref{newton},   where the linear relation of 
Eq.~(\ref{viscouslaw}) is shown to hold.   The slope of these curves then 
provide values for $\mu$ which can be compared to the Enskog result from 
Eq.~(\ref{viscosityenskog});  the results are shown in Fig.~\ref{Viscosity}.

Unlike the loss rate and thermal conductivity, we find that the Enskog
theory underestimates the shear viscosity at lower temperatures.  However, the trend of decreasing
transport with increasing T is the same.  Even for elastic particles, Enskog
theory is not expected to work to arbitrarily high solid fractions; as
density increases, deviations from Enskog theory are expected.  As the
inelasticity of particles increases, velocity correlations increase, reducing
collisional momentum transport and lowering the viscosity.

\section{Conclusion}
\label{conclusion}

Volumetric driving of granular media leads to nearly stationary states that are amenable to comparison with kinetic theory, allowing us to test the six points
of kinetic theory listed in Sec.~\ref{intro}.  Given that many of our 
simulations involve coefficients of restitution that are not near 1,
the general level of agreement with kinetic theory is surprisingly good,
suggesting that continuum approaches to dissipative granular media are capable not only
of qualitative, but also quantitative descriptions of real systems.
We now discuss each of the six points in turn.

\emph{Single particle distribution functions are nearly Boltzmann}.  For all
forcing types, temperatures (coefficients of restitution), and densities, the single particle distribution functions are close to Boltzmann distributions (Fig.~\ref{oneparticledist}).  
The deviations from Gaussian (Fig.~\ref{ratio}) are consistent with but
much smaller than those seen in  experiments on thin ($< 1$ layer deep) oscillated granular media~\cite{olafsen98}.  In those experiments, deviations appear to be due to 
spontaneous spatial variations in temperature that are not taken into account
in the analysis~\cite{urbachpersonal}; such variations become large for cooling, unforced granular media.  The smaller deviations
exhibited in our simulations may represent smaller temperature fluctuations.
 
\emph{Particle velocities are correlated.}  Standard kinetic theory assumes
molecular chaos: particle velocities are
uncorrelated.  In our simulations, as in simulations of cooling granular media~\cite{orza98}, strong velocity correlations (Fig.~\ref{vllvpp}) with a characteristic 
vortex structure~\cite{bizon99a} develop.  In the steady state, these correlations extend 
over the entire cell.
Molecular chaos is not required for kinetic theory;  a closure for the kinetic 
equation is critical.  Similar considerations have lead van Noije and Ernst
to apply ring kinetic theory~\cite{ernst71}, which allows for correlations
in particle velocities,  to a granular system~\cite{vannoije98b}.  

\emph{Spatial correlations are stronger than predicted by the Carnahan and 
Starling relation (Eq.~(\ref{CandS}))}.  Even for simulations using the Boltzmann bath, in which
velocity correlations are removed, the effect of inelasticity is to increase
the amount of spatial correlation at $r=\sigma$ (Fig.~\ref{gofrinelastic}(b)).  In other words, particles
are more likely in the inelastic case to be close to one another than
elastic particles in the same thermodynamic state.  The size of this effect
is dependent on the inelasticity, but can be greater than $15\%$. 

\emph{The equation of state overestimates the collisional contribution to 
pressure because it ignores velocity correlations.}  The factor in the equation of state describing the contribution from collisions, $G_s(\nu)$, as calculated
from the measurement of pressure and the equation of state, is smaller for white
noise and acceleration forcings than that predicted
by the Carnahan and Starling relation (Eqs.~(\ref{CandS}) and ~(\ref{Gg})), denoted $G_{CS}(\nu)$ (Fig.~\ref{Ginelastic}). In turn, $G_{CS}(\nu)$ is smaller than the actual
$G(\nu) \equiv \nu g(\nu)$,  as discussed in the previous paragraph.   Because velocity
correlations were ignored in the derivation of the equation of state (\ref{state}), the
pressure due to collisions that $G$ describes is overestimated.  To some 
degree,  the increased positional correlation and increased velocity
correlation work against one another; the first increases the collision 
frequency, while the latter decreases it.  The net result is that the $G_s(\nu)$
from the pressure measurement is closer to Carnahan and Starling (\ref{CandS}) $G_{CS}(\nu)$ than if there
were only velocity correlations (Fig.~\ref{gofrinelastic}(a)).

\emph{Newton's stress law works well for low stress.}   Even at the highest
inelasticity, $e=0.7$, no deviations from a linear relation between stress
and strain rate (Eq.~(\ref{newtonslaw})) were observed (Fig.~\ref{newton}).
However, in order to keep the temperature isotropic, we have limited ourselves
to cases in which $v^2 < T$; many flows of interest are supersonic, with 
average velocities much larger than $\sqrt{T}$.  Because
$\kappa$ depends on $T$, and therefore on position, Fourier's heat law was not
tested in the same manner that Newton's viscosity law was.

\emph{Temperature loss rate, $\gamma$, and thermal conductivity, $\kappa$, are reduced by inelasticity, while shear viscosity, $\mu$, 
is predicted relatively well by Enskog theory.}  For increasing inelasticity or density
in homogeneously forced runs, velocity correlations also increase.  
As a consequence, the relative collision velocity decreases (Fig.~\ref{PvsType}), leading to a
reduction in the temperature loss rate due to inelastic collisions (Figs.~\ref{gammaTpic}(b) and ~\ref{gammanupic}(b)).  In the
most severe cases examined, once corrected for variations in $e$, this
deviation could be as high as $20\%$.    Inhomogeneously forced runs allowed
calculation of $\mu$, and, assuming Fourier's law, $\kappa$;  $\mu$ never
deviated from the prediction of Enskog theory by more than $15\%$ (Fig.~\ref{Viscosity}), while
$\kappa$ was found to be smaller than predicted by a factor of two for high
inelasticities (Fig.~\ref{Conductivity}).  This differential success suggests that velocity correlations,
which should be present in both cases, are not responsible for the large
reduction in $\kappa$.  Rather, the inelasticity itself may be the cause.
Enskog theory, applied to granular media, assumes that $e \approx 1$; the
values of $\kappa_0$ and $\mu_0$ are the same as those for elastic particles.
When grains collide, energy is dissipated, so that the amount transported
collisionally is necessarily reduced as $e$ decreases.  On the other hand,
 momentum is still conserved, so that $\mu$ is relatively unaffected.  
Also,some deviation from Enskog theory is possible due to a term 
in the heat flux proportional to density gradients.

The results described above suggest a number of avenues for future research.
First, measurements of viscosity should be extended into the supersonic
regime.  Second, more extensive calculations of thermal conductivity at
different densities, and with different spatial forcings, should be undertaken
to ascertain the role of density gradients.  Third, time-dependent 
calculations should be performed, allowing measurement of the bulk viscosity.
Such calculations should also provide measurements of the frequency dependence
of the transport coefficients, which may be relevant for oscillated granular
media. Fourth, the granular continuum equations can be used to perform 
stability calculations on problems such as vertically oscillated granular
media~\cite{mythesis,bizon99b}. 
Finally, new forcing geometries should be explored, allowing 
direct comparison between particle simulations, continuum theories, and
experiments. 

\section{Acknowledgments}
We deeply thank Professor Jim Jenkins for helping us penetrate granular kinetic theory.  This work was supported by the Department of Energy Office of Basic Energy Sciences.

\bibliography{/u9new/bizon/grain/write/bib/sand}

\begin{thebibliography}{10}

\bibitem{campbell90}
C.~S. Campbell, Annu. Rev. Fluid Mech. {\bf 2},  57  (1990).

\bibitem{jaeger96}
H.~M. Jaeger, S.~R. Nagel, and R.~P. Behringer, Physics Today {\bf 49},  32
  (1996).

\bibitem{melo94}
F. Melo, P. Umbanhowar, and H.~L. Swinney, Phys. Rev. Lett. {\bf 72},  172
  (1994).

\bibitem{debruyn98}
J.~R. de~Bruyn, C. Bizon, M.~D. Shattuck, D. Goldman, J.~B. Swift, and H.~L.
  Swinney, Phys. Rev. Lett. {\bf 81},  1421  (1998).

\bibitem{lun84}
C.~K.~K. Lun, S.~B. Savage, D.~J. Jeffrey, and N. Chepurniy, J. Fluid Mech.
  {\bf 140},  223  (1983).

\bibitem{jenkins85}
J.~T. Jenkins and M.~W. Richman, Arch. Rat. Mech. Anal. {\bf 87},  355  (1985).

\bibitem{jenkins85a}
J.~T. Jenkins and M.~W. Richman, Phys. Fluids {\bf 28},  3485  (1985).

\bibitem{kadanoff97}
L.~P. Kadanoff, Rev. Mod. Phys. {\bf 71},  435  (1999).

\bibitem{bizon98}
C. Bizon, M.~D. Shattuck, J.~B. Swift, W.~D. McCormick, and H.~L. Swinney,
  Phys. Rev. Lett. {\bf 80},  57  (1998).

\bibitem{bizon98a}
C. Bizon, M.~D. Shattuck, J.~R. de~Bruyn, J.~B. Swift, W.~D. McCormick, and
  H.~L. Swinney, J. Stat. Phys. {\bf 93},  449  (1998), {C}olor pictures of
  granular convection can be found at
  http://chaos.ph.utexas.edu/errata/bizon98a.html.

\bibitem{olafsen98}
J.~S. Olafsen and J.~S. Urbach, Phys. Rev. Lett. {\bf 81},  4369  (1998).

\bibitem{delour98}
J. Delour, A. Kudrolli, and J.~P. Gollub, preprint  (1998).

\bibitem{oger96}
L. Oger, C. Annic, D. Bideau, R. Dai, and S.~B. Savage, J. Stat. Phys. {\bf
  82},  1047  (1996).

\bibitem{orza98}
J.~A.~G. Orza, R. Brito, T.~P.~C. van Noije, and M.~H. Ernst, Int. J. Mod.
  Phys. C {\bf 8},  953  (1998).

\bibitem{vannoije98}
T.~P.~C. van Noije, M.~H. Ernst, and R. Brito, Phys. Rev. E {\bf 57},  R4891
  (1998).

\bibitem{williams96}
D.~R.~M. Williams and F.~C. MacKintosh, Phys. Rev. E {\bf 54},  R9  (1996).

\bibitem{swift98}
M.~R. Swift, M. Boamf{\v{a}}, S.~J. Cornell, and A. Maritan, Phys. Rev. Lett.
  {\bf 80},  4410  (1998).

\bibitem{puglisi98}
A. Puglisi, V. Loreto, U.~M.~B. Marconi, A. Petri, and A. Vulpini, Phys. Rev.
  Lett. {\bf 81},  3848  (1998).

\bibitem{chapman}
S. Chapman and T.~G. Cowling, {\em The Mathematical Theory of Non-uniform
  Gases} (Cambridge University Press, London, 1970).

\bibitem{carnahan69}
N.~F. Carnahan and K.~E. Starling, J. Chem. Phys. {\bf 51},  635  (1969).

\bibitem{alder62}
B.~J. Alder and T.~E. Wainwright, Phys. Rev. {\bf 127},  359  (1962).

\bibitem{gass70}
D.~M. Gass, J. Chem. Phys. {\bf 54},  1898  (1970).

\bibitem{lubachevsky91}
B.~D. Lubachevsky, J. Comp. Phys. {\bf 94},  255  (1991).

\bibitem{marin93}
M. Mar{\'{\i}}n, D. Risso, and P. Cordero, J. Comput. Phys. {\bf 109},  306
  (1993).

\bibitem{lun87}
C.~K.~K. Lun, J. Appl. Mech. {\bf 54},  47  (1987).

\bibitem{lun91}
C.~K.~K. Lun, J. Fluid Mech. {\bf 233},  539  (1991).

\bibitem{mcnamara92}
S. McNamara and W.~R. Young, Phys. Fluids A {\bf 4},  496  (1992).

\bibitem{mcnamara94}
S. McNamara and W.~R. Young, Phys. Rev. E {\bf 50},  R28  (1994).

\bibitem{evans}
D.~J. Evans and G.~P. Morriss, {\em Statistical Mechanics of Nonequilibrium
  Liquids} (Academic Press, San Diego, 1990).

\bibitem{ippolito95}
I. Ippolito, C. Annic, J. Lema{\^i}tre, L. Oger, and D. Bideau, Phys. Rev. E
  {\bf 52},  2072  (1995).

\bibitem{hirschfelder}
J.~O. Hirschfelder, C.~F. Curtiss, and R.~B. Byrd, {\em Molecular Theory of
  Gases and Liquids} (Wiley, New York, 1954).

\bibitem{rapaportsbook}
D.~C. Rapaport, {\em The Art of Molecular Dynamics Simulation} (Cambridge
  University Press, Cambridge, 1980).

\bibitem{bizon99a}
C. Bizon, M.~D. Shattuck, J.~B. Swift, and H.~L. Swinney,  in {\em Dynamics:
  Models and Kinetic Methods for Nonequilibrium Many-Body S ystems (to
  appear)}, {\em NATO ASI Series E: Applied Sciences}, edited by J. Karkheck
  (Kluwer Academic publishers, Dordrecht, 1999).

\bibitem{brey97a}
J.~J. Brey, J.~W. Duffy, and A. Santos, J. Stat. Phys. {\bf 87},  1051  (1997).

\bibitem{urbachpersonal}
J.~S. Urbach, personal Communication.

\bibitem{ernst71}
M.~H. Ernst, E.~H. Hauge, and J.~M.~J. van Leeuwen, Phys. Rev. A {\bf 4},  2055
   (1971).

\bibitem{vannoije98b}
T.~P.~C. van Noije, M.~H. Ernst, and R. Brito, Physica A {\bf 251},  266
  (1998).

\bibitem{mythesis}
C. Bizon, Ph.D. thesis, University of Texas at Austin, 1998.

\bibitem{bizon99b}
C. Bizon, M.~D. Shattuck, and J.~B. Swift, To be published  (1999).

\end{thebibliography}
\bibliographystyle{myprsty}

\end{document}